\documentclass[fleqn,twoside]{article}
\usepackage{espcrc2}

\usepackage{graphicx}
\usepackage[figuresright]{rotating}

\newcommand{\AmS}{{\protect\the\textfont2
  A\kern-.1667em\lower.5ex\hbox{M}\kern-.125emS}}

\title{Phenomenology of charmless hadronic $B$ decays\thanks{Talk presented 
at the Sixth International Conference on Hyperons, 
Charm and Beauty Hadrons, IIT, Chicago, June 27--July 3 2004.}}

\author{Denis A. Suprun
\address{Enrico Fermi Institute and Department of Physics, 
University of Chicago, Chicago, IL 60637}$^,$\address{High Energy Theory Group, Brookhaven National Laboratory, Upton, NY 11973}}%

\begin{document}

\def \bea{\begin{eqnarray}}
\def \beq{\begin{equation}}
\def \eea{\end{eqnarray}}
\def \eeq{\end{equation}}

\begin{abstract}
The decays of $B$ mesons to a pair of charmless pseudoscalar mesons ($PP$ decays) or to a vector and pseudoscalar meson ($VP$ decays) have been analyzed within the framework of flavor SU(3) symmetry and the Kobayashi-Maskawa mechanism of $CP$ violation. Separate $PP$ and $VP$ fits proved to be successful in describing the experimental data (branching ratios, $CP$ asymmetries and time-dependent parameters). Decay magnitudes and relative weak and strong phases have been extracted from the fits. Values of the weak phase $\gamma$ were found to be consistent with the current indirect bounds from other analyses of CKM parameters.
\vspace{1pc}
\end{abstract}

\maketitle

\section{INTRODUCTION}

\begin{figure}[h]
\centerline{\includegraphics[width=.43\textwidth]{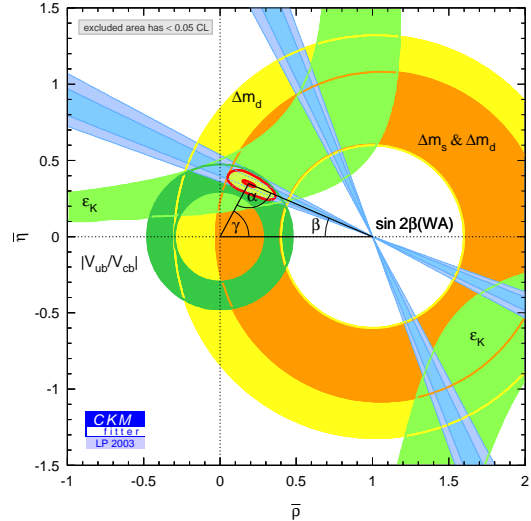}}
\caption{Current constraints on the Wolfenstein parameters $\rho$ and $\eta$~\cite{Hocker:2001xe}.
\label{fig:CKMFitter}}
\end{figure}

The main idea behind the study of $B$ meson decays is to get precise information on Cabibbo-Kobayashi-Maskawa (CKM) matrix elements. Testing the Kobayashi-Maskawa mechanism~\cite{Kobayashi:fv} of $CP$ violation in flavor physics requires many measurements of branching ratios and $CP$-violating observables. To consistently compare different results on common ground it is convenient to express them in terms of constraints on the apex of the CKM triangle in the $\rho-\eta$ plane (Fig.~\ref{fig:CKMFitter}) where $\rho$ and $\eta$ are parameters of the Wolfenstein parametrization of the CKM matrix. The weak phase $\beta \equiv \arg( -V_{cd}V_{cb}^*/V_{td}V_{tb}^*)$ was measured in $b \to c \bar c s$ decays (including $B^0 \to J/\psi K_S$~\cite{Aubert:2002ic}) with high precision. Currently, this CKM angle is determined to lie within a $5.8^{\circ}$ interval, $20.2^\circ \le \beta \le 26.0^\circ$, at 95\% confidence level~\cite{Hocker:2001xe}. Only indirect constraints exist for the other two CKM angles, 
$\alpha \equiv \arg(-V_{td}V_{tb}^*/V_{ud}V_{ub}^*)$ and 
$\gamma \equiv \arg(-V_{ud}V_{ub}^*/V_{cd}V_{cb}^*)$, with much larger  allowed ranges: $77^\circ \le \alpha \le 120^\circ$ and 
$39^\circ \le \gamma \le 80^\circ$ at 95\% confidence level.

Decays to two-body hadronic charmless final states are particularly useful since many of them involve more than one significant quark subprocess. Typically the weak phase difference between tree and penguin-type diagrams is equal to $\gamma$. When the strong phases are substantially different, too, a decay has the potential for displaying direct $CP$ asymmetries which can be observed in experiment. A reliable extraction of $\gamma$ is dependent on our ability to understand the pattern of strong phases in as wide as possible a set of decays. Final state interaction (FSI) strong phases involve nonperturbative long-distance physics and cannot be computed from first principles. A data-driven flavor topology approach based on the assumption of the SU(3) flavor symmetry~\cite{Chiang:2003pm,Chiang:2004nm,Suprun:2004} offers a way to extract FSI strong phases associated with individual topological amplitudes together with the weak phase $\gamma$ and topological decay amplitudes. 

In this analysis, we take flavor SU(3) symmetry 
\cite{DZ,Savage:1989ub,Chau:1990ay,Gronau:1994rj,Gronau:1995hn,%
Gronau:1995ng,Grinstein:1996us} 
as a working hypothesis. Assuming factorization, we take account of SU(3) symmetry breaking effects due to decay constant differences only when relating strangeness-conserving and strange-changing tree amplitudes. 
We do not expect factorization to work in penguin and color-suppressed amplitudes so we don't make any specific assumptions beyond the strict SU(3) symmetry. The ratios of strangeness-conserving and strange-changing  amplitudes for these two types of amplitudes are assumed to be completely determined by the ratio of the weak CKM matrix elements involved in either transition. 

From the results of two separate fits to $VP$ and $PP$ data one can extract information about fit parameters (decay amplitudes and their strong and weak phases), compare with other known constraints, and make predictions for  as-yet-unseen decay modes. The $VP$ analysis has particularly good  sensitivity to the CKM phase $\gamma$. This is driven in part by the pattern of tree-penguin interference in a wide variety of hadronic $B$ decays, and in part by the incorporation of time-dependent information on $B^0 \to \rho^\pm \pi^\mp$. The importance of $\rho^\pm \pi^\mp$ decays is not surprising as they were shown to be particularly sensitive to the CKM weak phase $\alpha$~\cite{Charles:2004jd,Gronau:2004tm}.
The values of $\gamma$ that are obtained in $VP$ and $PP$ fits are consistent with each other and with the current indirect bounds~\cite{Hocker:2001xe}. 

\section{$VP$ DECAYS}

The analysis of $VP$ decays~\cite{Chiang:2003pm} is based on measurements performed by the BaBar, Belle and CLEO collaborations on branching ratios, $CP$ asymmetries and/or time-dependent parameters in strangeness-preserving $\rho\pi$, $\omega\pi$, $\rho\eta$, and $\rho\eta'$ decays, and strangeness-changing $K^*\pi$, $K^*\eta$, $\rho K$, $\omega K$, and $\phi K$ decays. The total number of available data points is 34, including some quantities that do not affect the fit such as the time-dependent mixing-induced and direct asymmetries in the $\phi K_S$ decay, $S_{\phi K_S}$ and $A_{\phi K_S}$, and the $CP$ asymmetry $A_{CP}(B^+ \to \phi K^+)$. 

\begin{figure}[h]
\includegraphics[width=.484\textwidth]{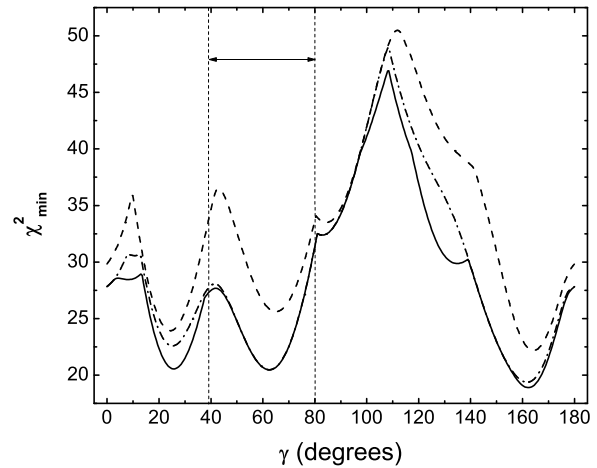}
\caption{$(\chi^2)_{\rm min}$, obtained by minimizing over all remaining fit parameters, as a function of the weak phase $\gamma$.  Dashed curve: $p'_V/p'_P = -1$ (24 d.o.f.); dash-dotted curve: $p'_V/p'_P$ real (23 d.o.f.); solid curve: $p'_V/p'_P$ complex (22 d.o.f.). Vertical dashed lines show the boundaries of the favored 95\% confidence level range of $\gamma$ ($39^{\circ} - 80^{\circ}$) from fits to CKM parameters~\cite{Hocker:2001xe} based on other measurements.
\label{fig:gc}}
\end{figure}

Plots of $\chi^2$ as a function of $\gamma$ for three version of $VP$ fits are shown in Fig.~\ref{fig:gc}. Three local minima are found, around $\gamma = 26^\circ$, $63^\circ$, and $162^\circ$. The solid line represents the fit 
with no constraints on the ratio $p'_V/p'_P$ of two different QCD penguin amplitudes in which the spectator quark hadronizes inside either a vector or a pseudoscalar final state meson. 

Only one of the three solutions, $\gamma=(63\pm6)^{\circ}$, is consistent with the current indirect bounds. It is also the only one that features small relative strong phase between two types of tree diagrams, $t_V$ and $t_P$. This feature is consistent with the predictions based on QCD factorization~\cite{Beneke:2003zv}. With $\chi^2=20.5/22$, the solution represents a good description of the current data within the present experimental uncertainties.

\section{$PP$ DECAYS}

A similar analysis of $PP$ decays~\cite{Chiang:2004nm} is based on measurements of branching ratios, $CP$ asymmetries and/or time-dependent parameters in strangeness-preserving $\pi\pi$, $\pi\eta$, and $\pi\eta'$ decays, and strangeness-changing $K \pi$, $\eta K$, and $\eta' K$ decays. The total number of $PP$ data points is 26. 

Several important lessons were learned while searching for a good $PP$ fit to the current data. First of all, a large relative strong phase 
$\delta_C\simeq-100^{\circ}$ between the color-suppressed $C$ and tree $T$ amplitudes is crucial for getting a satisfactory agreement between fit expectations and the experimental data. 
Although the SU(3) fit to $\pi\pi$, $\pi K$ decays 
(Fit I of Ref.~\cite{Chiang:2004nm}) is able to accommodate
the data when nontrivial $\delta_C$ is added to the fit, it prefers such values for $|C|$ and $|T|$ amplitudes that $|C/T|\simeq1.4$. Two effects are responsible for this unusually large amplitude ratio. One is the presence of final-state interactions, the other is the importance of a penguin amplitude $P_{tu}$ associated with intermediate $t$ and $u$ quarks. This penguin term features the same weak factors as tree-type amplitudes $T$ and $C$. 
When it is not explicitly taken into account as a fit parameter, it disguises itself as a part of the tree and color-suppressed amplitudes, interfering destructively with the former and constructively with the latter. 
When $P_{tu}$ is added as a fit parameter, the fit to $\pi\pi$, $\pi K$ data
(Fit II) separates $P_{tu}$ and tree-level amplitudes to predict a more reasonable $|C/T|\simeq0.5$ which is still larger than expected.

The values of both the $|C/T|$ amplitude ratio and the relative phase $\delta_C$ are roughly consistent with the result for the $C/T$ ratio inferred from $D\pi$ decays~\cite{Cheng:2001sc,Cheng:2004ru}. The extraction of this ratio from charmless $PP$ decays yields a larger $|C/T|$ ratio and a larger phase than expected from the QCD factorization approach. This indicates that soft final-state interactions play an important role in $B$ physics despite the naive expectation that products of energetic $B$ decays move away too fast to experience final-state rescattering.

\begin{figure}[t]
\includegraphics[width=.484\textwidth]{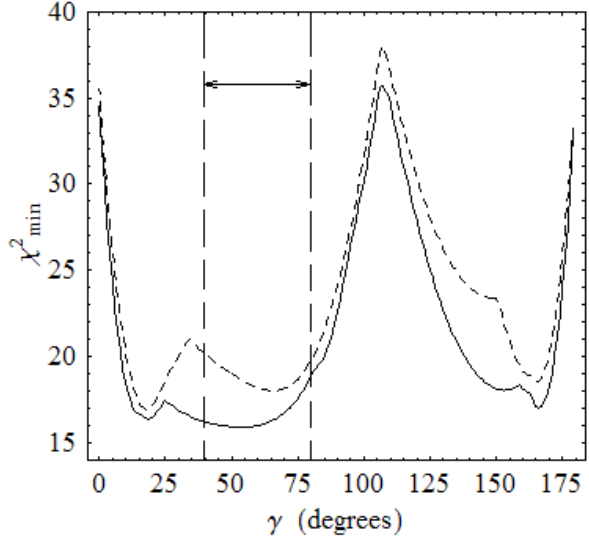}
\caption{$(\chi^2)_{\rm min}$, obtained by minimizing over all remaining fit
parameters, as a function of the weak phase $\gamma$. Dashed curve: Fit III of Ref.~\cite{Chiang:2004nm}; solid curve: Fit IV. Vertical dashed lines show the boundaries of the favored 95\% confidence level range of $\gamma$ ($39^{\circ} - 80^{\circ}$) from fits to CKM parameters~\cite{Hocker:2001xe} based on other measurements.
\label{fig:PPfit}}
\end{figure}

Fig.~\ref{fig:PPfit} shows plots of $\chi^2$ as a function of $\gamma$ for two versions of fits to {\it all} $PP$ data, including final states that involve flavor-singlet $\eta$ and $\eta'$ mesons. One of the fits (Fit IV) uses $S_{tu}$, a singlet-penguin amplitude associated with intermediate $t$ and $u$ quarks, as a fit parameter, the other neglects it. 
Both versions of $PP$ fits have a local $\chi^2$ minimum in the range $39^\circ \le \gamma \le 80^\circ$ allowed by global fits to phases of the CKM matrix~\cite{Hocker:2001xe}: $\gamma=(66^{+12}_{-16})^{\circ}$ in Fit III and $\gamma=(54^{+18}_{-24})^{\circ}$ in Fit IV.
The variation of central values of $\gamma$ between two fits is about 
$12^\circ$, providing an estimate of the systematic error associated with this topological approach.

\section{CONCLUSIONS AND NEW DEVELOPMENTS}

The decays of $B$ mesons to a pair of charmless mesons have been analyzed within a framework of flavor SU(3) symmetry of the topological quark diagrammatic approach. Acceptable separate fits to $PP$ and $VP$ branching ratios and $CP$ asymmetries were obtained with
tree, color-suppressed, penguin, and electroweak penguin amplitudes. The penguin amplitude $P_{tu}$ associated with intermediate $t$ and $u$ quarks was found to considerably improve the quality of $PP$ fits. Contrary to expectations, the value of relative strong phase $\delta_C$ between $C$ and $T$ amplitudes and the value of the $|C/T|$ ratio were found to be large, hinting at the presence of final-state interaction effects. So far, the data is accommodated well within the SM; the largest deviation from fit predictions does not exceed $1.7\sigma$.

New experimental data on $K \pi$ decays~\cite{Aubert:2004dn} 
made the so-called ``$K \pi$ puzzle" less severe.  $R_c\equiv2\Gamma(K^+\pi^0)/\Gamma(K^0\pi^+)$ and $R_n\equiv\Gamma(K^+\pi^-)/2\Gamma(K^0\pi^0)$ are expected to be equal in the limit of small color-suppressed amplitudes. The new data determine the difference between $R_c$ and $R_n$ to be approximately equal to $0.21\pm0.13$. The discrepancy is under $2\sigma$, a smaller significance than before.

A joint fit to all data on charmless hadronic $B$ decays is currently being developed with the weak phase $\gamma$ as a common parameter for the $PP$ and $VP$ sectors of the fit~\cite{Suprun:2004}. Just as in the case of separate $PP$ and $VP$ fits, one can extract the magnitudes and relative phases of different topological amplitudes and make predictions for rates and $CP$ asymmetries in as-yet-unseen decay modes, including $B_s$ decays. Preliminary results of the joint fit are roughly consistent with those obtained in the  analyses of $B \to VP$ and $B \to PP$ decays. The global minimum of $\chi^2$ is achieved at the weak phase $\gamma \simeq 55^{\circ}$. It favors $\gamma$ within the range $51^\circ$--$59^\circ$ at the $1 \sigma$ level, and 48$^\circ$--62$^\circ$ at 95\% confidence level.

\medskip
I am grateful to Cheng-Wei Chiang, Michael Gronau, Zumin Luo, and  Jonathan Rosner for enjoyable collaborations.


\begin{thebibliography}{99}

\bibitem{Kobayashi:fv}
M.~Kobayashi and T.~Maskawa,
Prog.\ Theor.\ Phys.\  {\bf 49}, 652 (1973).

\bibitem{Aubert:2002ic}
BABAR Collaboration, B.~Aubert {\it et al.},
Phys.\ Rev.\ Lett.\  {\bf 89}, 201802 (2002);
Belle Collaboration, K.~Abe {\it et al.},
Phys.\ Rev.\ D {\bf 66}, 071102 (2002).
Updated results may be found on the web site
{\tt http://www.slac.stanford.edu/xorg/ \\
hfag/triangle/index.html}. 

\bibitem{Hocker:2001xe}
A.~Hocker, H.~Lacker, S.~Laplace and F.~Le Diberder,
Eur.\ Phys.\ J.\ C {\bf 21}, 225 (2001). 
Updated results may be found on the web site
{\tt http://ckmfitter.in2p3.fr}. 

\bibitem{Chiang:2003pm}
C.~W.~Chiang, M.~Gronau, Z.~Luo, J.~L.~Rosner and D.~A.~Suprun,
Phys.\ Rev.\ D {\bf 69}, 034001 (2004).

\bibitem{Chiang:2004nm}
C.~W.~Chiang, M.~Gronau, J.~L.~Rosner and D.~A.~Suprun,
Phys.\ Rev.\ D {\bf 70}, 034020 (2004).

\bibitem{Suprun:2004}
D.~A.~Suprun, 
in progress.

\bibitem{DZ} D. Zeppenfeld, Zeit.\ Phys.\ C {\bf 8}, 77 (1981).

\bibitem{Savage:1989ub} 
M. Savage and M. Wise, 
Phys.\ Rev.\ D {\bf 39}, 3346 (1989); {\it ibid.} {\bf 40}, 3127(E) (1989).

\bibitem{Chau:1990ay} 
L. L. Chau {\it et al.}, 
Phys.\ Rev.\ D {\bf 43}, 2176 (1991); {\it ibid.} {\bf 58}, 019902 (1998).

\bibitem{Gronau:1994rj}
M.~Gronau, O.~F.~Hernandez, D.~London and J.~L.~Rosner,
Phys.\ Rev.\ D {\bf 50}, 4529 (1994). 
 
\bibitem{Gronau:1995hn}
M.~Gronau, O.~F.~Hernandez, D.~London and J.~L.~Rosner,
Phys.\ Rev.\ D {\bf 52}, 6374 (1995). 

\bibitem{Gronau:1995ng}
M.~Gronau and J.~L.~Rosner,
Phys.\ Rev.\ D {\bf 53}, 2516 (1996).

\bibitem{Grinstein:1996us}
B.~Grinstein and R.~F.~Lebed,
Phys.\ Rev.\ D {\bf 53}, 6344 (1996).

\bibitem{Charles:2004jd}
CKMfitter Group Collaboration, J.~Charles {\it et al.},
arXiv:hep-ph/0406184.

\bibitem{Gronau:2004tm}
M.~Gronau and J.~Zupan,
arXiv:hep-ph/0407002.

\bibitem{Beneke:2003zv}
M.~Beneke and M.~Neubert,
Nucl.\ Phys.\ B {\bf 675}, 333 (2003).

\bibitem{Cheng:2001sc}
H.~Y.~Cheng,
Phys.\ Rev.\ D {\bf 65}, 094012 (2002).

\bibitem{Cheng:2004ru}
H.~Y.~Cheng, C.~K.~Chua and A.~Soni,
arXiv:hep-ph/0409317.

\bibitem{Aubert:2004dn}
BABAR Collaboration, B.~Aubert {\it et al.},
arXiv:hep-ex/0408062; hep-ex/0408080; hep-ex/0408081.


\end{thebibliography}
\end{document}